\documentclass[12pt]{article}  
\usepackage[dvips]{graphicx}
\usepackage{amsmath,amssymb,amsbsy}
\usepackage{amsfonts}
\usepackage{epsfig}
\newcommand{\beq}{ \begin{equation}}
\newcommand{\eeq}{\end{equation}}
\newcommand{\ba}{\begin{array}}

\newcommand{\beqa}{\begin{eqnarray}}
\newcommand{\eeqa}{\end{eqnarray}}
\newcommand{\bear}{\begin{array}{c}}
\newcommand{\bearr}{\begin{array}{cc}}
\newcommand{\ear}{\end{array}}

\newcommand{\R}{I\kern-.3em{R}}
\newcommand{\no}{\nonumber}

\begin{document}
\thispagestyle{empty}
\begin{center}

{\Large\bf{
Statistical approach of pion\\
\vskip 0.3cm
parton distributions from Drell-Yan process}}
\vskip1.4cm
{\bf Claude Bourrely}
\vskip 0.3cm
Aix Marseille Univ, Univ Toulon, CNRS, CPT, Marseille, France
\vskip 0.5cm
{\bf Jacques Soffer}
\vskip 0.3cm
Physics Department, Temple University,\\
1835 N, 12th Street, Philadelphia, PA 19122-6082, USA
\vskip 0.5cm
{\bf Abstract}\end{center}

The quantum statistical approach proposed more than one decade ago was used to determine the parton distributions for the proton by considering a large set of accurate Deep Inelastic Scattering experimental results. We propose to extend this work to extract the parton distributions for the pion by using data on lepton pair production from various experiments. This global next-to-leading order QCD analysis leads to a good description of several Drell-Yan $\pi^- W$ data. The resulting parton distributions are compared with earlier determinations. We will also discuss the difference between nucleon and pion structure in the same approach.

\vskip 0.5cm


\vskip 0.5cm

\newpage
\section{\small {Introduction}}
Deep Inelastic Scattering (DIS) of leptons and nucleons is indeed our main source of information to
study the internal nucleon structure in terms of parton distributions. 
Several years ago a new set of parton distribution functions (PDF) was constructed in the
framework of a statistical approach of the nucleon \cite{bbs1}. For quarks
(antiquarks), the building blocks are the helicity dependent distributions
$q^{\pm}(x)$ ($\bar q^{\pm}(x)$). This allows to describe simultaneously the
unpolarized distributions $q(x)= q^{+}(x)+q^{-}(x)$ and the helicity
distributions $\Delta q(x) = q^{+}(x)-q^{-}(x)$ (similarly for antiquarks). At
the initial energy scale $Q_0^2$, these distributions
are expressed in terms of a quasi Fermi-Dirac function. The flavor asymmetry for the light sea, {\it i.e.} $\bar d (x) > \bar
u (x)$, observed in the data is built in. This is simply understood in terms
of the Pauli exclusion principle, based on the fact that the proton contains
two up-quarks and only one down-quark. The chiral properties of QCD lead to
strong relations between $q(x)$ and $\bar q (x)$.
Concerning the gluon distribution $G(x,Q_0^2)$ it is given in
terms of a quasi Bose-Einstein function, with only {\it one free parameter}.
The predictive power of this approach lies partly in the DIS sector, but mainly in the rich domain of
hadronic collisions, up to LHC energies, as reported in Refs.~\cite{bbs2,bbs5,bbsW,bs3}.\\
As mesons are not available as targets for performing DIS
experiments, the existing experimental inputs to extract the meson PDFs come almost exclusively from Drell-Yan
process and direct photon production with meson beams. The Drell-Yan process does not involve fragmentation functions, which is a main advantage to extract PDFs. The Drell-Yan process describes massive lepton pair production and it dominates the production of $W^{\pm}$ and $Z$ gauge bosons
in $pp$ and $\bar pp$  collisions. The statistical approach was used successfully to describe these reactions up to LHC energies \cite{bbps1, bbps2} and our goal in this paper is to
extend this framework to study the meson structure using data from Drell-Yan dimuon production by charged pions beams on nuclear targets \cite{E615,NA10}. 
The structure of the pion which has zero spin is obviously much simpler than the nucleon structure which carries spin 1/2. For the nucleon one can define eight quark distributions with transverse momentum $k_T$ and only three (unpolarized, helicity, transversity) survive after $k_T$ integration \cite{bacchetta}. For the pion in addition to the unpolarized, one can only define one transversely polarized distribution with $k_T$, which vanishes after $k_T$ integration \cite{diehl}. Consequently, as we will see, the {\it direct} extension of the statistical approach to the pseudoscalar mesons is not possible, because one cannot define  the helicity dependent distributions $q^{\pm}(x)$ ($\bar q^{\pm}(x)$) inside a pion.

The paper is organized as follows. In Section 2, we give the main points of
our approach to construct the $\pi^{\pm}$ parton distributions for the pion. In Section 3 the Drell-Yan cross section with $\pi^{\pm}$ beams on nuclear targets is given and we describe our method to determine the free parameters 
of the PDFs with the set of experimental data we have used. In Section 4, we show and discuss the results obtained from the analysis and the $\pi^{\pm}$ distributions. We give our final remarks and conclusions  in the last section.
\newpage
\section{\small{The $\pi^{\pm}$ distributions $U, D, \bar U, \bar D$ in the statistical approach}}
The pion $\pi^+$ is made of two constituent quarks $u$ and $\bar d$ and the corresponding distributions will be denoted by $U$ and $D$. 
In addition we have the two corresponding sea quarks those distributions are denoted by $\bar U$ and $\bar D$.\\
For $\pi^-$ which is made of two constituent quarks $d$ and $\bar u$ and two sea quarks, we don't need new distributions, because
from charge symmetry we have
\beq
  U= u_{\pi^+}=\bar u_{\pi^-}, D=\bar d_{\pi^+}=d_ {\pi^-}, \bar U=\bar u_{\pi^+}=u_{\pi^-}, \bar D=d_{\pi^+}=\bar d_{\pi-}.
  \label{eq1}
\eeq
In the general method used to construct the quark distributions in the nucleon case $xq(x)$, as recalled in the introduction, we used a sum of two terms $xq(x)= xq^{+}(x)+xq^{-}(x)$.
Following this general procedure, we propose, in the meson case, to parametrize the quark distributions also as a sum of two terms with similar expressions. Clearly in this case one cannot associate these two terms to helicity distributions. Therefore for the
quark distributions $Q=U,D$, we have $xQ(x) = x Q^+(x) + x Q^-(x)$ where \footnote{The diffractive term introduced in the nucleon case is important in the low $x$ region ($x < 0.1$). However since the available DY data does not reach this region, it will be omitted in the pion case}
\beq
  xQ^{\pm}(x) = \frac{A_{Q}X_Q^{\pm}x^{b_{Q}}}{\mbox{exp}[(x - X_Q^{\pm})/\bar x] + 1} ,
  \label{eq2}
\eeq
at the input energy scale  $Q_0^2=1 \mbox{GeV}^2$. 
 Note that so far we have introduced for each flavor {\it four} parameters,
$A_Q, b_Q$ and {\it two thermodynamical potentials} $X_Q^{\pm}$. The value of the {\it universal temperature} $\bar x$ will be taken as in the nucleon case \cite{bs3}, namely $\bar x = 0.090$ .
Concerning the antiquarks distributions $\bar Q =\bar U, \bar D$, similarly we have  $x\bar Q(x) = x\bar Q^+(x) + x\bar Q^-(x)$ where
\beq
 x\bar Q^{\pm}(x) = \frac{\bar A (X_Q^{\mp})^{-1}x^{\bar b}}{\mbox{exp}[(x + X_Q^{\mp})/\bar x] + 1} .
 \label{eq3}
\eeq 
 We have used  exactly the same rule, as in the nucleon case, to relate quarks and antiquarks and we have introduced {\it two} additional parameter $ \bar A$ and $\bar b$.\\
To summarize the meson PDFs in the light quark sector are parametrized in terms of  {\it ten} parameters, a situation very similar to that of the nucleon case. However only {\it eight} are free parameters since we have {\it two} normalization constrains, namely
\beq
 \int_{0}^{1} (U(x) - \bar U(x) ) dx = 1~~Ê\mbox{and}~~ \int_{0}^{1} (D(x) - \bar D(x) ) dx = 1.
 \label{eq4}
 \eeq
Finally the gluon distribution will be parametrized, at the input energy scale  $Q_0^2=1 \mbox{GeV}^2$, as in the nucleon case $xG_{\pi}(x) = {A_G} x^{b_G}/(\mbox{exp}(x/\bar x) - 1)$.
$A_G$ is determined by the momentum sum rule, which reads,
\beq
 \int_{0}^{1}x[ (U(x) + \bar U(x) ) + D(x) + \bar D(x) + G_{\pi}(x) ] dx = 1,
 \label{eq5}
\eeq
and $b_G$ which controls the small $x$ behavior will be taken the value obtained in the nucleon case \cite{bs3}, namely $b_G = 1.020$.\\
These meson distributions will be obtained after the determination of the parameters from a next-to-leading (NLO) QCD fit of two sets
of Drell-Yan dimuon production by $\pi^-$ beams on tungsten targets \cite{E615,NA10}.

\section{\small {The Drell-Yan cross section with $\pi^{\pm}$ beams on nuclear targets}}

The Drell-Yan cross section $\pi p$ collisions at NLO is written as
\beqa
 \frac{d\sigma}{dM^2dx_F} & =&  \frac{4\pi \alpha^2}{9M^2s}\sum_i e_i^2 
\int_{x_1}^1 dt_1\int_{x_2}^1 dt_2 \no \\  
&& \left\{ \left[ \frac{d^{DY} \sigma}{dM^2dx_F}+ \frac{d^{A} 
\sigma}{dM^2dx_F}\right]
\cdot [Q_i^{\pi} (t_1)\cdot \bar q_i^p (t_2) 
+ \bar Q_i^{\pi} (t_1)\cdot  q_i^p (t_2)] \right. \no \\
&&  +\frac{d^{C} \sigma}{dM^2dx_F}\left\{G_{\pi}(t_1)[q_i^p(t_2)+
\bar q_i^p(t_2)] + G_{p}(t_2)[Q_i^{\pi}(t_1)+\bar Q_i^{\pi}(t_1)] \right\}\,,
\label{eq6}
\eeqa
where $d\sigma^{DY}, d\sigma^{A}, d\sigma^{C}$ represent the contributions leading-order (LO),
 annihilation and Compton, they are defined in Ref. \cite{sutton}. In the above formula $M^2= x_1x_2s$ and $\tau = M^2/s$, 
 $Q_i^{\pi} = U, D$ are the $\pi^{\pm}$ distributions and $q_i^p$ are 
the proton distributions and $x_F = x_1 - x_2$, we use the scale $Q^2= M^2$.

Since we want to describe the E615 and NA10 data, which are given in terms of $\tau$, $x_F$ and the energy $s$, we need to express this cross section in terms of these three
variables. We have at LO, which is the dominant term,
\beq
 \frac{d^2\sigma}{d\sqrt{\tau}dx_F} = \frac{8\pi \alpha^2}{9s\sqrt{\tau}(x_1 + x_2)} \sum_i e_i^2[Q_i ^{\pi}(x_1)\cdot \bar q_i^p (x_2) + \bar Q_i^{\pi} (x_1)\cdot q_i^p (x_2)] .
 \label{eq7} 
 \eeq
  Note that $ x_1 = 1/2[ x_F + \sqrt{x_F^2 + 4\tau}]$ and $ x_2 = 1/2[ -x_F + \sqrt{x_F^2 + 4\tau}]$.\\
 If we consider Drell-Yan in  $\pi^+ p$ collisions  the $\sum_i e_i^2$ term reads
\beq
 1/9[4 U(x_1)\bar u_p(x_2) + \bar D(x_1) \bar d_p(x_2) + 4 \bar U(x_1) u_p(x_2) + D(x_1) d_p(x_2)] . 
 \label{eq8}
\eeq
The last two terms being obtained from the first two terms, by exchanging quarks and antiquarks.
Note that the last term is largely dominant because it is quark-quark, whereas the other three terms are either quark-sea or sea-sea.\\
Let us now turn to Drell-Yan in $\pi^- p$ collisions and in this case one has using charge symmetry
\beq
 1/9[4 \bar U(x_1)\bar u_p(x_2) + D(x_1) \bar d_p(x_2) + 4 U(x_1) u_p(x_2) + \bar D (x_1) d_p(x_2)] , 
 \label{eq9}
\eeq
which is deduced from $\pi^+ p$ by the substitutions $ U \leftrightarrows \bar U$ and $ D \leftrightarrows \bar D$  .\\
Now the dominant term is the third one because it is quark-quark, whereas the other three terms are either quark-sea or sea-sea.\\
We will consider DY data from $\pi^{\pm} A$ collisions, where A denotes a nuclei of atomic mass A with Z protons and A-Z neutrons. So in addition
to $\pi^{\pm} p$ collisions, we need to include $\pi^{\pm} n$ collisions. Once more it can be directly obtained from $\pi^{\pm} p$ by using charge symmetry
which relates proton and neutron, namely $u_p = d_n$, $d_p = u_n$ and similarly for the corresponding antiquarks.\\
So for Drell-Yan in $\pi^- n$ collisions, one gets

\beq
 1/9[4 \bar U(x_1)\bar d_p(x_2) + D(x_1) \bar u_p(x_2) + 4 U(x_1) d_p(x_2) + \bar D (x_1) u_p(x_2)] , 
 \label{eq10}
\eeq
and, finally for $\pi^- A$ collisions, one obtains
\beqa
 && 1/9[4 \bar U(x_1)[(Z/A) \bar u_p(x_2) + (1 - Z/A)\bar d_p(x_2)] + D(x_1)[(Z/A) \bar d_p(x_2) + (1 - Z/A) \bar u_p(x_2)] ]
\no \\
 && + [quark   \leftrightarrows  antiquark].
 \label{eq11}
\eeqa
To complete the expression of the cross section for $\pi^- A$ collisions, similar procedure must be done for the NLO contributions in Eq.(\ref{eq6}).

\section{\small{Results from the analysis and $\pi^{\pm}$ distributions}}
In order to determine the parameters introduced in Section 2, we have performed a NLO QCD global fitting procedure using two sets of Drell-Yan dimuon production by $\pi^-$ beams on tungsten targets, with dimuon mass between the $J/\Psi$ and $\Upsilon$ resonances and $0 <  x_F < 0.85$.
 We present in the table the details of the number of points and corresponding $\chi^2$ for each experiment, with a total of 117 data points for a total $\chi^2$ of 103.\\
 \begin{table}[hbp]
\begin{center}
\begin{tabular}{ c c c }
\hline
process &$\chi^2$ &nb points \\
\hline\raisebox{0pt}[12pt][6pt]
{$d \sigma$ ($\pi^-~W$) E615 \cite{E615}}       & 68    & 73  \\[4pt]
{$d \sigma$  ($\pi^-~W$) NA10 \cite{NA10} }   & 35    & 44  \\[4pt]
\hline\raisebox{0pt}[12pt][6pt]
{Total }                                 & 103  & 117  \\[4pt]
\hline
\end{tabular}
\caption {Detailed $\chi^2$ for the DY cross sections.}
\label{table1}
\end{center}
\end{table}
The PDF QCD evolution was done using of the HOPPET program \cite{hoppet}, the minimization of the $\chi^2$ was performed by the
CERN MINUIT program \cite{minuit} and we obtained the following parameters:
\begin{eqnarray}
\nonumber
A_U = 0.537  \pm 0.100,~ A_D = 0.346 \pm  0.050,\\ 
b_U = 0.048 \pm 0.001, ~ b_D = 0.466 \pm 0.014,
\label{eq12}
\end{eqnarray}
and four potentials 
\begin{eqnarray}
\nonumber
X_{U}^+= 0.787 \pm 0.007,~X_{U}^-= 0.185 \pm 0.030,\\ X_{D}^+= 0.866 \pm 0.024,~ X_{D}^-= 0.718\pm 0.044.
\label{eq13}
\end{eqnarray}
In addition we found
\beq
 \bar A = 1.706\pm 0.080,~~Ê\bar b = 0.157 \pm 0.010,~~ A_G =  30.111  \pm 0.680.
\label{eq14}
\eeq

\begin{figure}[hbp]
\begin{center}
\includegraphics[width=9.5cm]{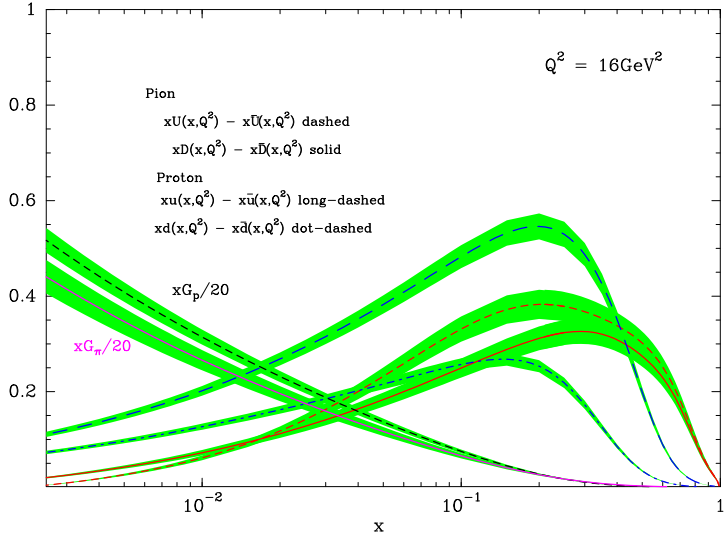}
\end{center}
\caption[*]{\baselineskip 1pt
The different parton distributions for the pion $xU(x,Q^2) - x\bar U(x,Q^2)$,   $xD(x,Q^2) - x\bar D(x,Q^2)$ and $xG_{\pi}(x,Q^2)$ versus $x$, after NLO QCD evolution at $Q^2=16\mbox{GeV}^2$. They are compared to the corresponding parton distributions for the nucleon, $xu(x,Q^2) - x\bar u(x,Q^2)$,   $xd(x,Q^2) - x\bar d(x,Q^2)$ and $xG_{p}(x,Q^2)$. The error bands for the pion distributions are consistent with Eqs. (\ref{eq13},\ref{eq14}) and for the nucleon distributions are those of Ref. \cite{bs3}.}
\label{PDF}
\end{figure}
\begin{figure}[hbp]
\begin{center}
\includegraphics[width=10.5cm]{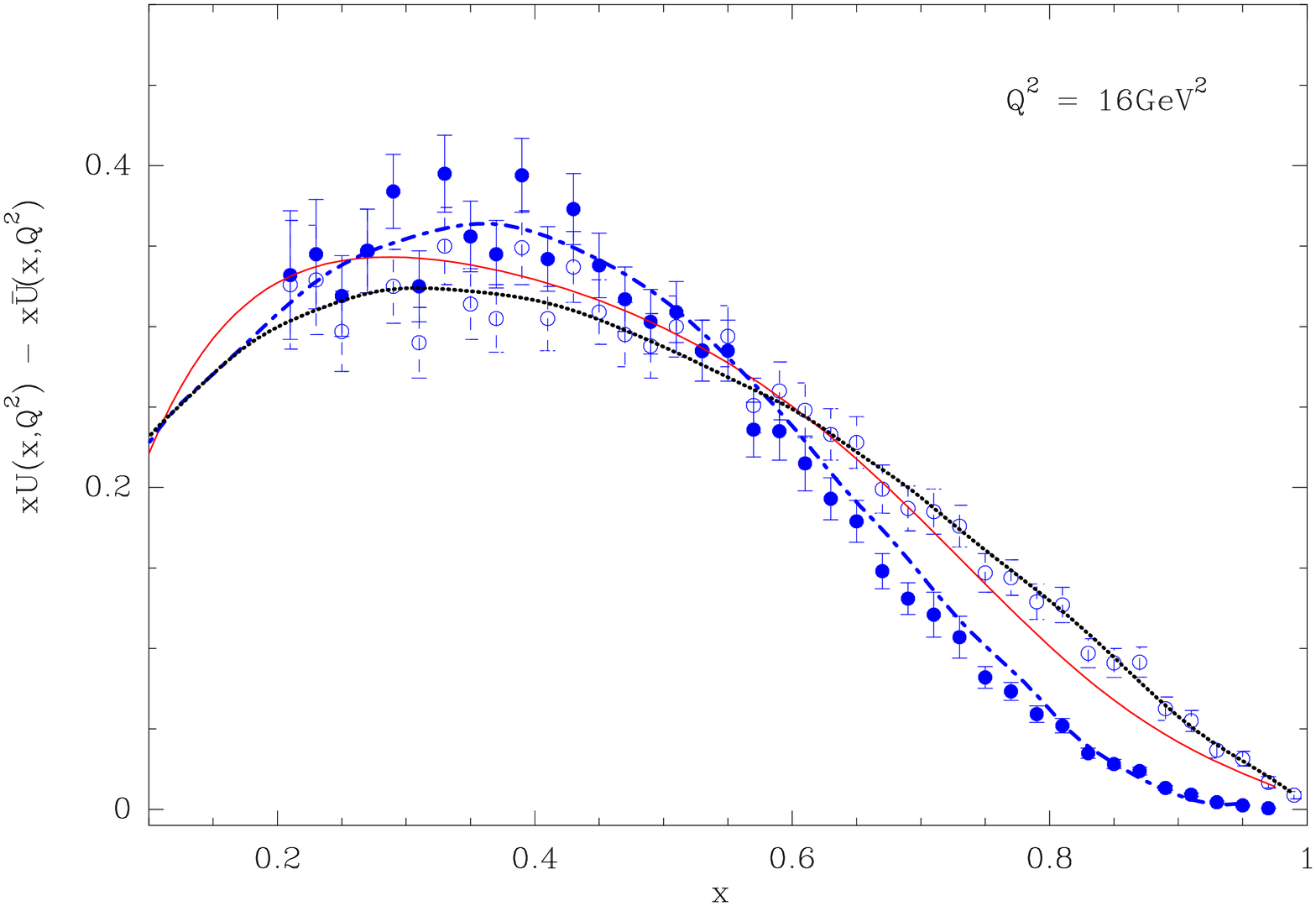}
\caption[*]{\baselineskip 1pt
The parton distribution for the pion $xU(x,Q^2) - x\bar U(x,Q^2)$ (solid curve) versus $x$, after NLO QCD evolution at $Q^2=16\mbox{GeV}^2$ compared to
the determination from Ref.\cite{sutton} (dotted curve) and from Ref.\cite{werner1} (dashed-dotted curve). For $x > 0.4$ it coincides with the nonperturbative Dyson-Schwinger equation Ref.\cite{roberts} approach. The white circles are data from Ref.\cite{E615} and the black circles
are the rescaling data from Ref.\cite{chen}.}
\label{Fpi}
\end{center}
\end{figure}
\begin{figure}[hb]
\begin{center}
\includegraphics[width=9.5cm]{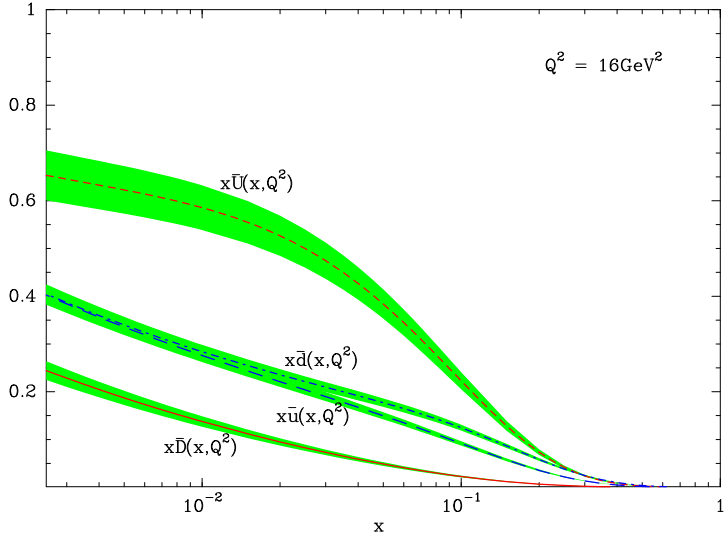}
\caption[*]{\baselineskip 1pt
The antiquark distributions for the pion $x\bar U(x,Q^2)$,  $x\bar D(x,Q^2)$ versus $x$, after NLO QCD evolution at $Q^2=16\mbox{GeV}^2$. They are compared to the corresponding antiquark distributions for the nucleon, $ x\bar u(x,Q^2)$,   $x\bar d(x,Q^2)$. The error bands for the pion distributions are consistent with Eqs. (\ref{eq13},\ref{eq14}) and for the nucleon distributions are those of Ref. \cite{bs3}.}
\label{sea}
\end{center}
\end{figure}
\begin{figure}[hb]
\begin{center}
\includegraphics[width=8.5cm]{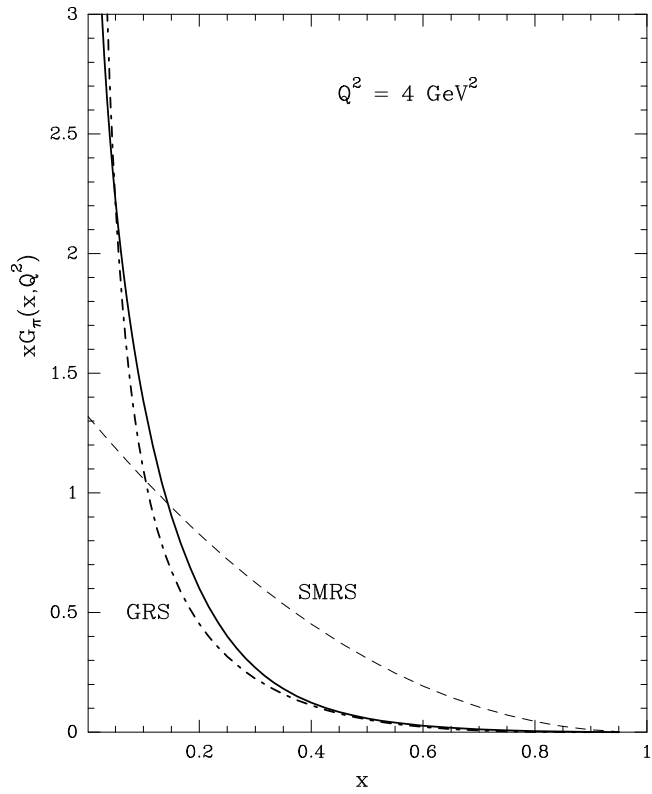}
\caption[*]{\baselineskip 1pt
Our gluon distribution for the pion $xG(x,Q^2)$ versus $x$, at $Q^2=4\mbox{GeV}^2$ (solid curve), compared to
the determinations from Ref.\cite{GRS} (dotted curve) and from Ref.\cite{sutton} (dashed-dashed curve).}
\label{gluon4}
\end{center}
\end{figure}

We display in Fig.\ref{PDF} the two valence distributions of the pion $xQ(x,Q^2) - x\bar Q(x,Q^2)$ ($Q = U,D$)  
and the gluon distribution $xG_{\pi}(x,Q^2)$, versus $x$ after NLO QCD evolution at $Q^2=16\mbox{GeV}^2$.
We see that for $x < 0.6$ both valence distributions do not coincide, at variance with the extraction from Ref.\cite{E615}, which was assuming their equality as a starting point. It would be interesting to have an accurate experimental determination in the region $x < 0.6$. The corresponding distributions for the nucleon obtained in Ref.\cite{bs3} are also shown and it is clear that they both have a much faster falloff in the high-$x$ region, compared to that of the meson. This high-$x$ behavior is compared in Fig. \ref{Fpi} with other theoretical determinations and also with the E615 data \cite{E615}, well described by the first attempt to perform a NLO QCD analysis of the data done in Ref.\cite{sutton}, making some simplifying assumptions like the equality of the two valence distributions and a $SU(3)$ symmetric pion sea. However it leads to considerably harder valence distribution at high $x$,
than predicted by perturbative QCD counting rules \cite{crule} and nonperturbative Dyson-Schwinger equation \cite{roberts} approaches. By considering soft-gluon resummation \cite{werner1,werner2}, as shown in Ref.\cite{chen}, the E615 data \cite{E615} must be rescaled \footnote{Prof. C.D. Roberts kindly provided the numerical values displayed in Fig.\ref{Fpi}} and the linear falloff $(1 - x)$ turns to a softer behavior $(1 - x)^2$. The statistical valence which, at the moment, does not take into account the soft-gluon resummation
corrections, lies between these two behaviors. It is worth noting that the nucleon PDFs are different in all these determinations of the PDFs of the pion.\\
The gluon distribution in the pion is rather similar to that of the nucleon as shown in Fig.\ref{PDF} and as expected,
they both largely dominate in the low-$x$ region and they both carry approximately 50{\%} of the hadron momentum.\\
Similarly the different antiquark distributions $x\bar Q(x,Q^2)$ ($\bar Q = \bar U,\bar D$), versus $x$, after NLO QCD evolution at $Q^2=16\mbox{GeV}^2$ are shown in Fig.\ref{sea}. The corresponding distributions for the nucleon obtained in Ref.\cite{bs3} are also shown with our long-standing prediction $x\bar d(x,Q^2) > x\bar u(x,Q^2)$  \cite{bbs1,bs4}, which remains to be confirmed by the SeaQuest experiment \cite{reimer}. In the pion case, the situation is reverse and, as expected, we find $x\bar U(x,Q^2) > x\bar D(x,Q^2)$. The fact that $x\bar U(x,Q^2)$ largely dominates in the very low-$x$ region is an interesting prediction.\\
 Clearly it is relevant to compare pion and nucleon distributions and in Ref. \cite{GRS} they are related by using a constituant quark model, which allows also to be applied to the kaon
 meson, worth mentioning. For illustration we display in Fig. \ref{gluon4} a comparison between our pion gluon distribution with those of Ref. \cite{sutton} and Ref. \cite{GRS}. There is a rather good agreement between our gluon distribution and that of Ref. \cite{GRS} but both disagree with that of Ref. \cite{sutton}, which has a rather slow $x$ fall off.\\
Let us now discuss an important point of our analysis. We have assumed charge symmetry, as indicated in Eq.\ref{eq1}, but not isospin symmetry, which would mean $U=D$ and 
$\bar U = \bar D$. So it is legitimate to ask what would be the consequences of this more restrictive assumption. Therefore we have made several tests but we were unable to get
such a good $\chi^2$, as reported above (see Table 1). The best we could do was obtained by allowing overall normalisation K-factors different from 1, as done in earlier analyses.
We conclude that to a certain extend and within our fitting procedure, the available data is demanding isospin symmetry breaking.


\begin{figure}[hbp]
\begin{center}
\includegraphics[height=13.0cm]{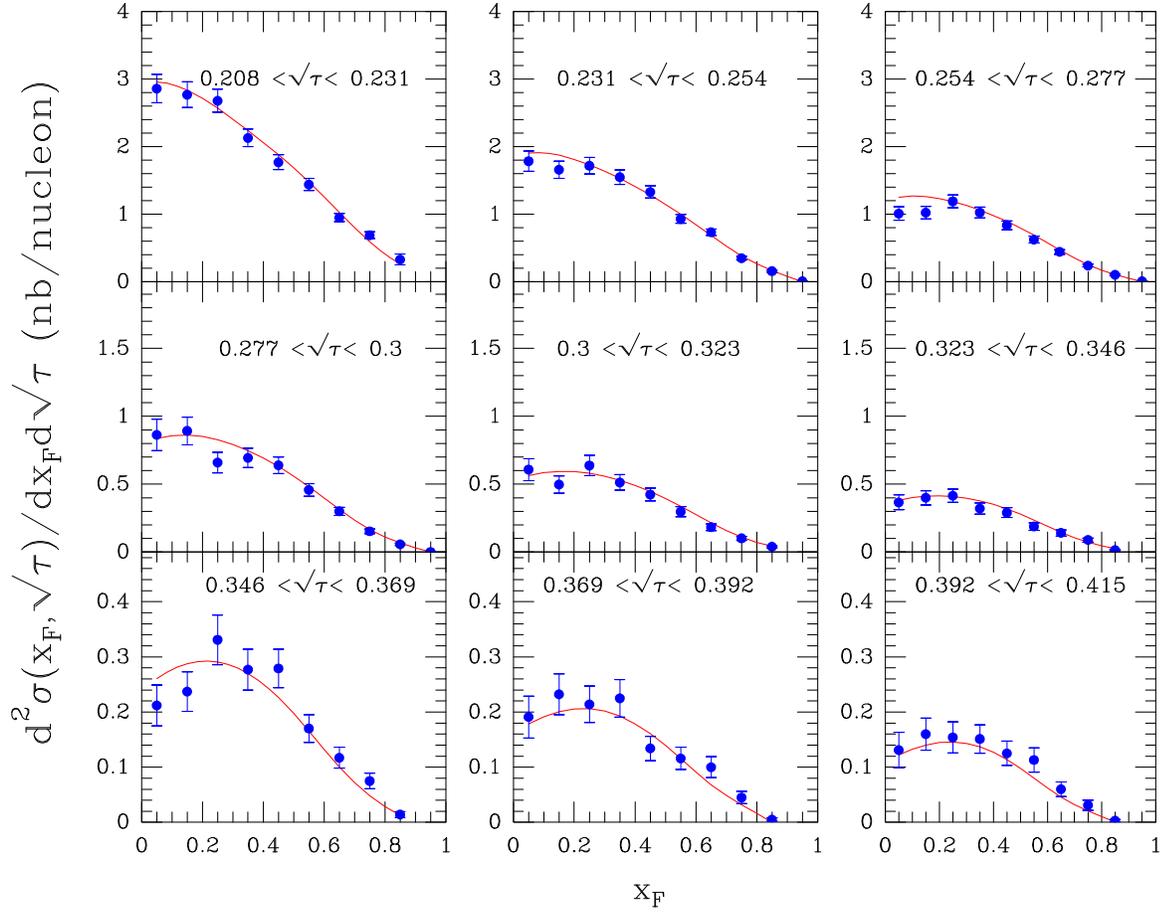}
\end{center}
\caption[*]{\baselineskip 1pt
Drell-Yan data from the E615 experiment $\pi^- W$ at $P_{lab}^{\pi} = 252 \mbox{GeV}$ \cite{E615}.  $d^2\sigma/d\sqrt{\tau}dx_F$ versus $x_F$ for several $\sqrt{\tau}$ intervals. The solid curves are the results of our global fit.}
\label{fi1}
\end{figure}

\begin{figure}[hbp]
\begin{center}
\includegraphics[height=13.0cm]{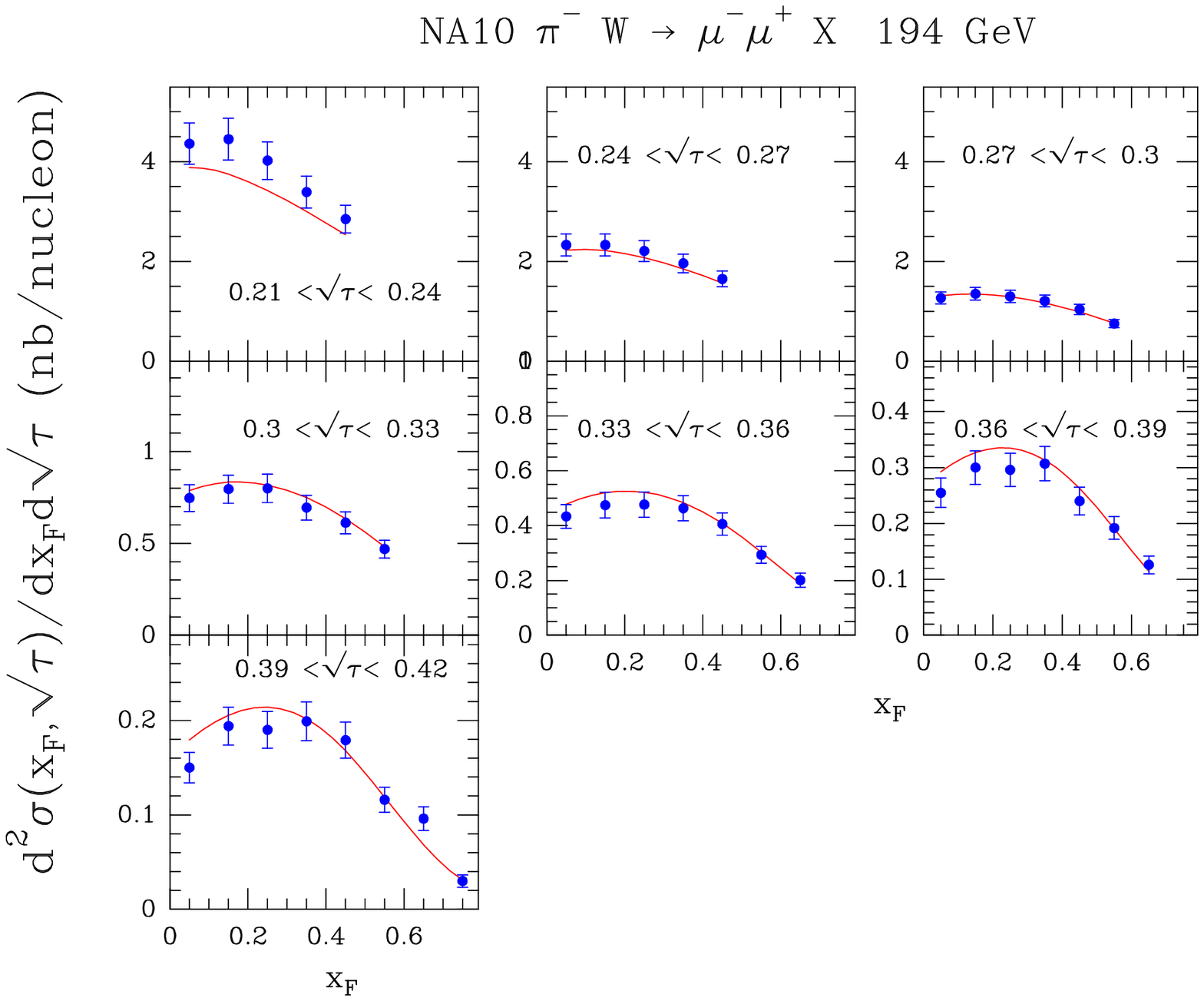}
\end{center}
\caption[*]{\baselineskip 1pt
Drell-Yan data from the NA10 experiment $\pi^- W$ at $P_{lab}^{\pi} = 194 \mbox{GeV}$ \cite{NA10}.  $d^2\sigma/d\sqrt{\tau}dx_F$ versus $x_F$ for several $\sqrt{\tau}$ intervals. The solid curves are the results of our global fit.}
\label{fi4}
\end{figure}

\begin{figure}[hbp]
\begin{center}
\includegraphics[height=13.0cm]{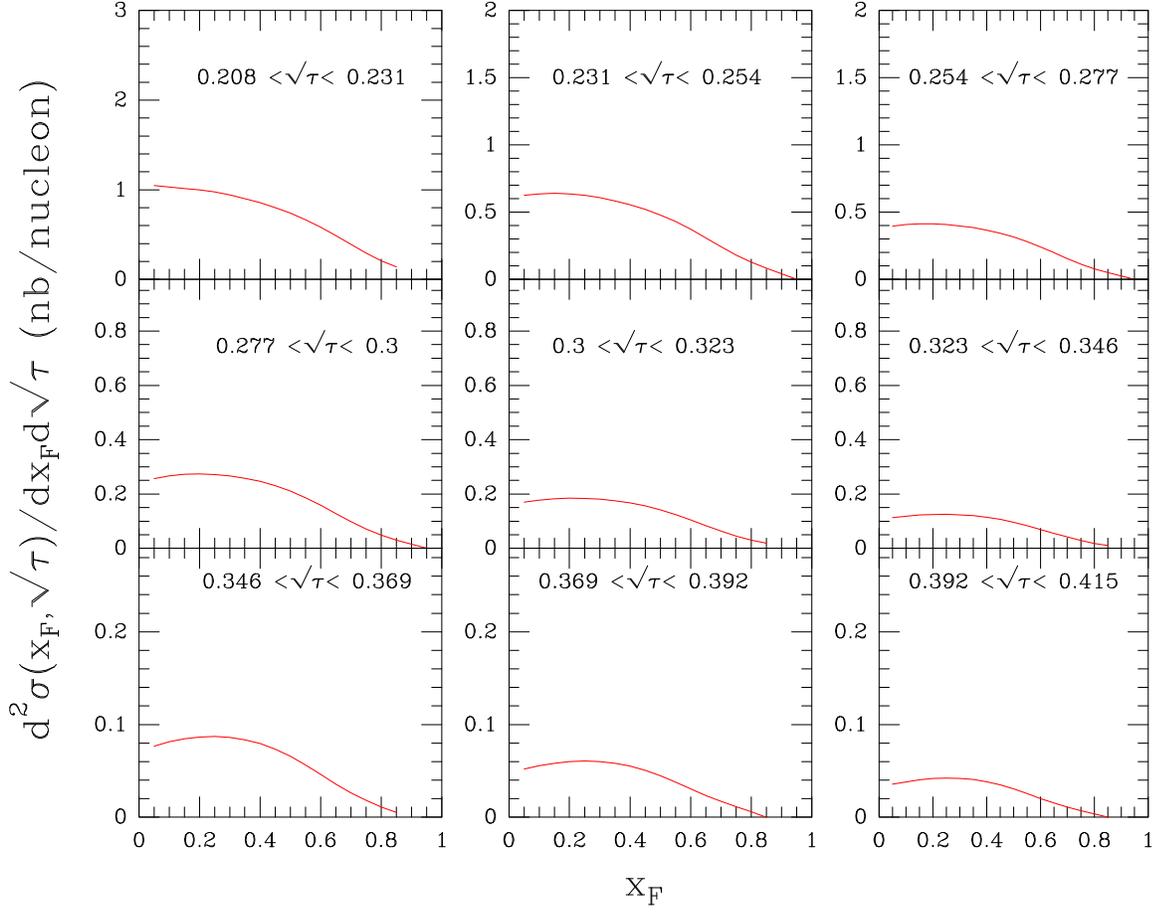}
\end{center}
\caption[*]{\baselineskip 1pt
Our predicted DY cross section $d^2\sigma/d\sqrt{\tau}dx_F$ versus $x_F$ for several $\sqrt{\tau}$ intervals for $\pi^+ W$ at $P_{lab}^{\pi} = 190 \mbox{GeV}$.}
\label{fi6}
\end{figure}

\begin{figure}[hbp]
\begin{center}
\includegraphics[height=12.0cm]{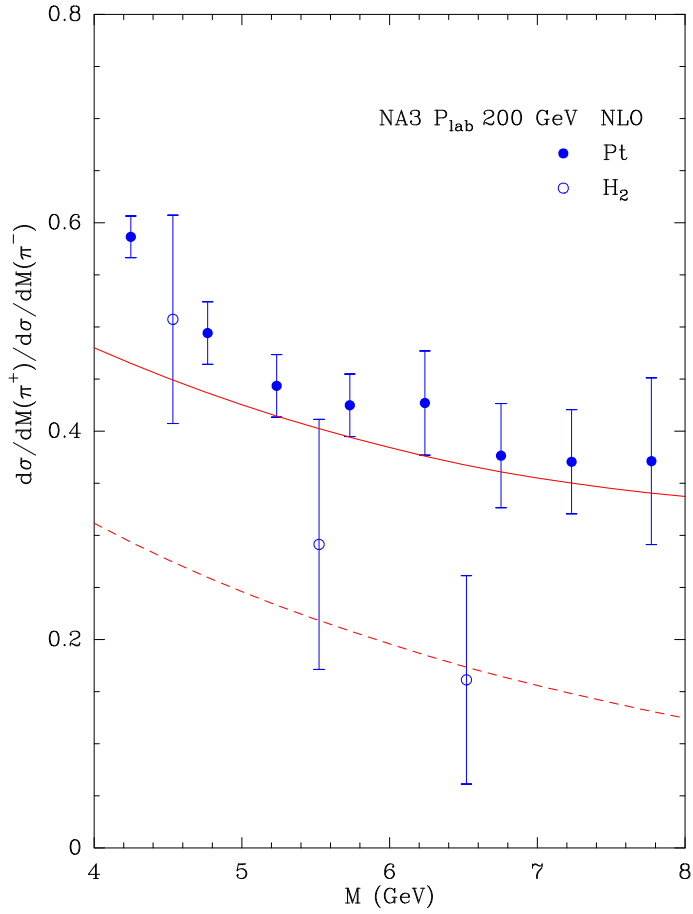}
\end{center}
\caption[*]{\baselineskip 1pt
Our predicted DY cross section ratio $d\sigma/dM(\pi^+)/d\sigma/dM(\pi^-)$ versus the dimuon mass M for 
$\pi^{\pm} Pt, H_2$ at $P_{lab}^{\pi} = 200 \mbox{GeV}$, compared to data Ref. \cite{callot}.}
\label{fi7}
\end{figure}

\clearpage
Let us now move to the description of the DY cross sections obtained by E615 and NA10, as displayed in Figs. \ref{fi1},\ref{fi4}, which show a remarkable agreement.
 We have noticed that the NLO corrections are important mainly in the large $x_F$ region ( $x_F > 0.3$), like the soft-gluon resummation effects mentioned above.\\
Finally another interesting piece of information to test our approach, comes from the use of $\pi^+$ beams. Given the scarcity of existing data, several new projects are under study and we give in Fig. \ref{fi6} our predicted DY cross section $d^2\sigma/d\sqrt{\tau}dx_F$ for $\pi^+ W$ versus $x_F$ for several $\sqrt{\tau}$ intervals, at $P_{lab}^{\pi} = 190 \mbox{GeV}$. By comparing with Fig. \ref{fi4}, one can check that, the cross section ratio $d^2\sigma/d\sqrt{\tau}dx_F(\pi^+)/d^2\sigma/d\sqrt{\tau}dx_F(\pi^-)$ is smaller than one and it decreases with increasing $\sqrt{\tau}$, at fixed $x_F$ and for fixed $\sqrt{\tau}$ it increases with $x_F$. We also show in Fig.\ref{fi7} the ratio $d\sigma/dM(\pi^+)/d\sigma/dM(\pi^-)$ versus the dimuon mass $M$ for $\pi^{\pm}$ on Platinium and Hydrogen targets at $P_{lab}^{\pi} = 200 \mbox{GeV}$, which is consistent with the data Ref. \cite{callot}.


\clearpage
\section{Concluding remarks}
The quantum statistical approach to describe the nucleon parton distributions, we proposed sixteen years ago, has been extended to extract the parton distributions for the pion by using lepton pair production from several DY $\pi^- W$ experiments. \\
We were able to separate valence and sea contributions in the light flavor sector with a small number of free parameters and we found that the best description of the data, within our fitting procedure, requires isospin symmetry breaking. These four distributions were compared to our nucleon PDFs and also to other determinations. We find that the gluon in the pion is very similar to the gluon in the nucleon. Due to the limited amount of available data at the moment, the extracted pion PDFs are not as precise as the nucleon PDFs. We made definite predictions for future $\pi^+ W$ experiments. One can also think of studying the kaon structure using kaon induced DY lepton pair production and some first attempts were done in Ref.\cite{chen,peng,GRS}. The DY lepton pair production process
with different beams is a rapid developing field with new perspectives to study hadron struture \cite{trento17}.\\

\begin{center}
{\bf Acknowledgments}\\
\end{center}

J.S is grateful to Prof. Jen Chieh Peng for suggesting this work, to Prof. Craig D. Roberts and Prof. Werner Vogelsang for very valuable comments.\\ We both thank Dr. Stephane Platchkov for providing useful information on existing cross section data and future DY experiments.

\end{document}